\documentclass[twocolumn,showpacs,preprintnumbers,amsmath,amssymb]{revtex4}

\usepackage{graphicx}
\usepackage{dcolumn}
\usepackage{bm}

\begin{document}


\boldmath
\title{Photoproduction of $\eta$-mesic $^3$He}
\unboldmath

\author{
  M. Pfeiffer$^1$,
  J.~Ahrens$^2$,
  J.R.M.~Annand$^3$,
  R.~Beck$^2$,
  G.~Caselotti$^2$,
  S.~Cherepnya$^7$,
  K.~F{\"o}hl$^5$,
  L.S.~Fog$^3$,
  D.~Hornidge$^2$,
  S.~Janssen$^1$,
  V.~Kashevarov$^7$,
  R.~Kondratiev$^{6}$,
  M.~Kotulla$^{4}$,
  B.~Krusche$^{4}$,
  J.C.~McGeorge$^3$,   
  I.J.D.~MacGregor$^3$,
  K.~Mengel$^1$,
  J.G.~Messchendorp$^8$,
  V.~Metag$^1$,
  R.~Novotny$^1$,
  M.~Rost$^2$,
  S.~Sack$^1$,
  R.~Sanderson$^3$,
  S.~Schadmand$^1$,
  A.~Thomas$^2$,
  D.P.~Watts$^3$
}
\affiliation{%
  $^1$II. Physikalisches Institut, Universit\"at Gie{\ss}en,
      D--35392 Gie{\ss}en, Germany\\
  $^2$Institut f\"ur Kernphysik, Johannes-Gutenberg-Universit\"at Mainz,
      D--55099 Mainz, Germany \\
  $^3$Department of Physics and Astronomy, University of Glasgow,
      Glasgow G128QQ, UK \\
  $^4$Department of Physics and Astronomy,
      University of Basel, CH-4056 Basel, Switzerland\\
  $^5$Department of Physics and Astronomy,
      University of Edinburgh, Edinburgh EH8 9YL, UK\\
  $^6$Institute for Nuclear Research, 60-th October Anniversary prospect 7a,
      117312, Moscow, Russia\\
  $^7$P.N. Lebedev Physical Institute, Leninsky Prospect 53, 117924, Moscow, Russia\\
  $^8$KVI, Zernikelaan 25, 9747 AA Groningen, The Netherlands
}%

\date{\today}

\begin{abstract}
The photoproduction of $\eta$-mesic $^3$He has
been investigated using the TAPS calorimeter at the Mainz Microtron accelerator
facility MAMI. The total inclusive cross section for the reaction
$\gamma$$^3$He$\rightarrow$$\eta$X has been measured for photon energies from
threshold to 820 MeV. The total and angular differential coherent $\eta$
cross sections have been extracted up to energies of 745 MeV. A resonance-like
structure just above the $\eta$ production threshold with an isotropic
angular distribution suggests the existence of a resonant quasi-bound state.
This is supported by studies of a competing decay channel of such a
quasi-bound $\eta$-mesic nucleus into $\pi^0$pX. A binding energy of
(-4.4$\pm$4.2)
MeV and a width of (25.6$\pm$6.1) MeV is deduced for the quasi-bound 
$\eta$-mesic state in $^3$He.

\end{abstract}

\boldmath
\maketitle
\unboldmath
It has long been discussed whether the attractive strong interaction among
mesons and nucleons may lead to the existence of mesic nuclei. Deeply bound
pionic states are known to exist due to the superposition of the attractive
electromagnetic and repulsive strong interaction. In the case of the neutral
$\eta$-meson an attractive strong $\eta$N interaction would allow for
the formation of $\eta$-mesic nuclei.
First predictions for such states
were based on investigations of the $\eta$-nucleon scattering
length $a_{\eta{}N}$. 
Its real part can be interpreted as a measure for the
scattering of the initial particles while its imaginary part accounts for
losses into other channels. 
First estimates of the $\eta$-N scattering length have been extracted from
coupled channel analyses, performed by Bhalerao and Liu \cite{bhalerao} in
1985. They found a 
scattering length $a_{\eta{}N}$ of (0.27+i0.22) fm. The corresponding phase
shifts have positive values which indicates an attractive potential. Based on
these calculations, Liu and Haider \cite{liu} claimed that bound states
between the $\eta$ and a nucleus should be possible for nuclei with mass number
A$>$10.\\
In 1991, Ueda \cite{ueda} predicted the existence of a quasibound $\eta$NN state with
I=0, J=1. 
These theoretical studies have raised hopes of experimentally discovering
quasi-bound states of $\eta$ mesons in light nuclei. If they exist, one
may expect them to be narrow in few-nucleon systems. The rapid slope of the
near-threshold amplitude in pd$\rightarrow$$\eta$$^3$He was interpreted by Wilkin
\cite{wilkin} as an 
indirect evidence for the existence of $\eta$-mesic nuclei. 
A recent experiment \cite{sokols} has studied the possible
photoproduction of $\eta$-mesic Boron via the reaction
$\gamma+^{12}C\rightarrow{}p+^{11}_\eta{}B\rightarrow\pi^++n+X$ with an
intermediate $S_{11}$(1535) resonance which decays into $\pi^+n$. The
correlated energies and momenta of pion and neutron have been used as a
signature for the reaction channel.\\
Concerning the lightest nuclei, photoproduction experiments on the deuteron
\cite{bernd_paper,weiss}
and on $^4$He 
\cite{hejny2} 
have shown modifications of the
near-threshold behaviour via a comparison with models based on an
impulse approximation. Sibirtsev et al. \cite{sibirtsev} took final
state interactions
of the N-N system as well as in the $\eta$-N system into account in their
calculation of the quasifree $\eta$ photoproduction off $^2$H. The
data were best reproduced for a scattering length $a_{\eta{}N}$=(0.42 +
i0.34) fm. Other analyses (e.g. \cite{wilkin,kaiser,nievess,wycech}) of the
scattering length yield varying results,
especially for the real part of the scattering length. The $\eta$-$^3$He
scattering length, extracted from photoproduction or hadroproduction, provides
information on the possibility of a bound state of $\eta$ and nucleus.
In this Letter, the results for coherent $\eta$ production off $^3$He and the 
$\eta$-mesic decay channel into $\pi^0$pX are presented in terms of a 
possible photoproduction of $\eta$-mesic $^3$He.\\
\begin{figure}
  \includegraphics[width=0.99\columnwidth]{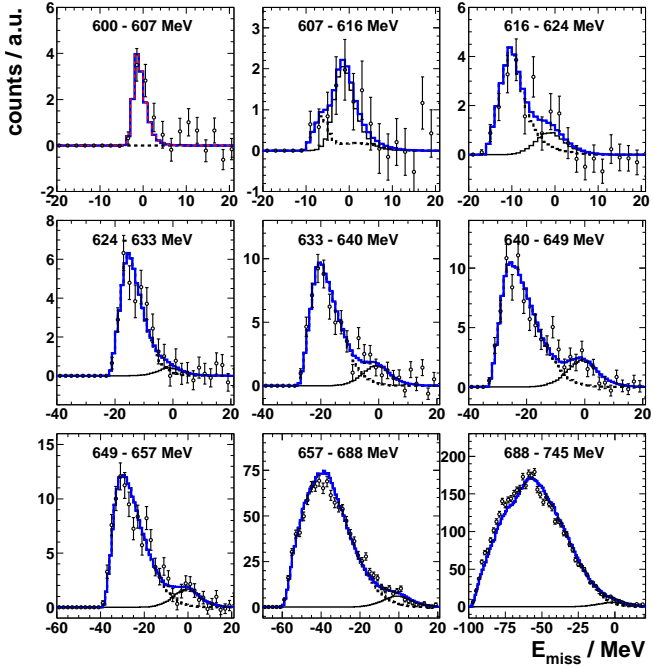}
  \caption{Missing energy plots
  assuming coherent reaction kinematics in different regimes of the incoming
  photon energy. The simulations of the coherent reaction (thin solid line)
and the quasifree reaction (dashed line) are adjusted to fit the data points.
The thick solid lines show the sum of both simulations. At threshold
(upper left plot, $E_\gamma$=600 - 606 MeV), the $\eta$ production proceeds
solely via the coherent process. In the other plots, the coherent cross section was derived from
the ratio of the quasifree and coherent part (see
text).} 
\label{miss_e}
\end{figure}
The experiment was performed at the 
electron accelerator facility Mainz Microtron (MAMI) 
\cite{ahrens94s}.
Photons were produced in a thin radiator foil and their energies
were determined using the Glasgow tagged photon spectrometer 
\cite{hall}.
The tagged photon energy range was 275-820 MeV with an energy resolution of
$\sim$ 2 MeV.
The reaction products were detected in the TAPS photon spectrometer
\cite{rainerieee,gabler}. TAPS
consisted of six blocks each with 62 hexagonally shaped BaF$_2$
scintillation detectors arranged in an 8x8 matrix and a forward wall with
138 BaF$_2$ detectors in an upright 11x14 arrangement. The length of the
crystals is 250 mm ($\sim$12 radiation lengths) with an inner
diameter of 59 mm plus a cylindrical end cap  
of 25 mm length (54 mm diameter). The crystals were read out by
photomultipliers. The six blocks were located in a horizontal plane around
the target point at angles of 153$^\circ$, 104$^\circ$, 55$^\circ$,
-54$^\circ$, -103$^\circ$, -152$^\circ$ with respect to the beam axis
at distances of 55-59 cm. The 
forward wall was positioned at 0$^\circ$ at a distance of 55 cm. The
solid angle covered by this setup is around 40\% of the full solid angle
\cite{kottu_paper}.
The target had an effective length of 115 mm and a diameter of 43 mm. It was
filled with liquid $^3$He. More detail can be found in \cite{tariq}.\\
\begin{figure}
   \includegraphics[width=0.95\columnwidth]{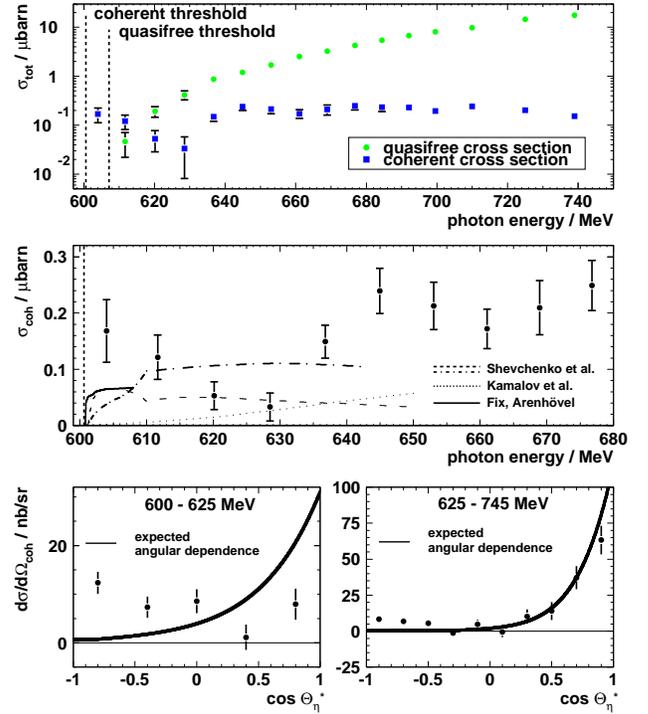}
  \caption[Coherent $\eta$ production]{Total cross sections for coherent
  and quasifree $\eta$ production off $^3$He and angular
 distributions in the $\gamma$-$^3$He CM system. The error bars contain statistical
and systematic uncertainties. The curves in the 
angular differential plots show the calculated coherent $\eta$ cross section
according to d$\sigma$/d$\Omega$$\sim$F$^2$(q$^2$) scaled to the measured
data. Calculations of the coherent cross section are compared to the data.
\label{cs_coh}}
\end{figure}
For the identification of the $\eta$ and $\pi^0$ mesons the double photon
decay channel was used. The two photon invariant mass was calculated from
the momenta of the measured photons. The experimental resolution
is 21 MeV (FWHM) for $\pi^0$ detection and 60 MeV for the $\eta$ meson.
The 
identification of protons is achieved via the characteristic relation between
deposited energy and time of flight, as well as pulse shape
analysis \cite{rainer96}. The calibration of the proton energy was performed by
calculating the missing mass for the reaction $\gamma$$^3$He$\rightarrow$$\eta$pd.
Energy losses of the protons and quenching effects in the
detector have been corrected for such that the missing mass corresponds to
that of the proton.
Random coincidences in the tagging spectrometer have been taken into account by
subtracting background events before and after the prompt time coincidence
peak.\\
The inclusive $\eta$ cross section was deduced from the rate of $\eta$ events
divided by the number of target nuclei per cm$^2$, the photon flux, the
branching ratio of $\eta$ into 2$\gamma$, and the detection and analysis
efficiency. These efficiencies were determined by Monte Carlo simulations
using the GEANT3 code. The photon flux was measured by counting the number
of scattered electrons in the tagger. 
The missing energy, E$_{miss}$, for the
coherent reaction $\gamma$$^3$He$\rightarrow$$\eta$$^3$He was obtained by
subtracting the calculated $\eta$ energy from two body kinematics from the
measured $\eta$ energy in the $\eta$$^3$He center of momentum system.
Coherent events should be centered at $E_{miss}$= 0 MeV
while break-up events are distributed at negative values of the missing
energy. 
\begin{figure}
 \begin{center}
   \includegraphics[width=0.43\textwidth]{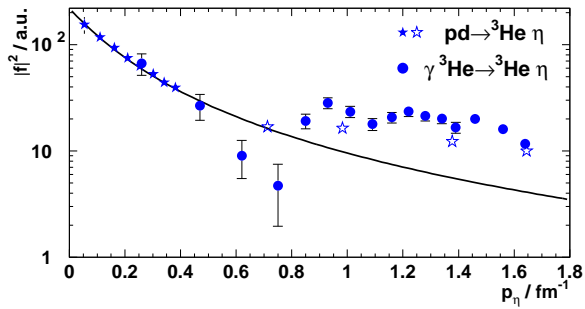}
  \caption{Average amplitude squared as a function of the CM $\eta$ momentum
for the proton induced reaction (full stars \cite{mayer}, open stars
\cite{bilger}) compared to the photoproduction data. The normalization of
the data sets is arbitrary. The solid curve is an 
optical model fit of the near-threshold data (figure adapted from
\cite{bilger}). 
\label{saclay}}
 \end{center}
\end{figure}
\begin{figure*}
   \includegraphics[width=0.745\textwidth]{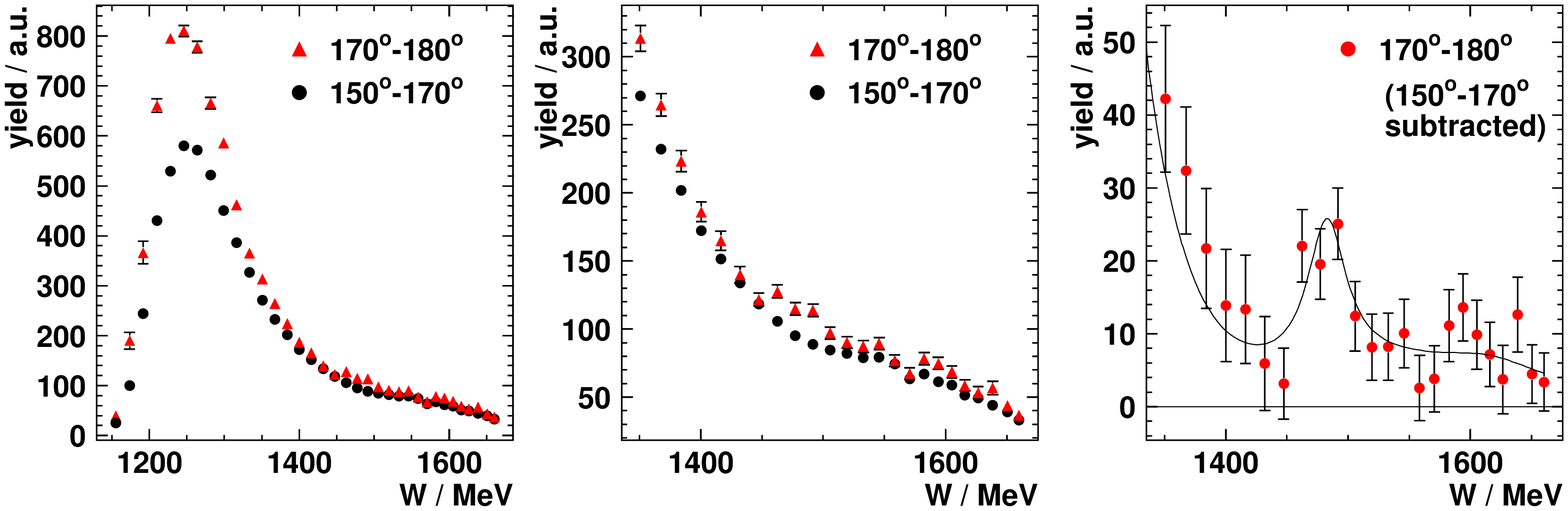}
  \caption{Left and center:
  Excitation function of the $\pi^0$-proton production for opening angles of
  170$^\circ$ - 180$^\circ$ (triangles) compared to opening angles of
  150$^\circ$ - 170$^\circ$ (circles) in the $\gamma$-$^3$He center of
  momentum system. Right: Difference of both distributions with a
  Breit-Wigner distribution plus background fitted to the data. 
\label{mesic}}
\end{figure*}
The fraction of
coherent $\eta$ production (thin solid line in fig. \ref{miss_e}) and
quasifree $\eta$ production (dashed line) was obtained by fitting
simulations for both production mechanisms to the experimental data. The
simulation for the quasifree production includes the elementary production
cross section on the proton folded with the Fermi momentum distribution.
The resulting coherent $\eta$
cross section is plotted in fig. \ref{cs_coh}. 
The peak-like behaviour of the near threshold cross section can be
interpreted as a 
first sign for a modified $\eta$ production process. To investigate this
behaviour further, the angular distribution for the two energy ranges
600-625 MeV and 625-745 
MeV are plotted in the lower part of fig. \ref{cs_coh}. While the
latter distribution agrees with the expected angular dependence
(d$\sigma$/d$\Omega$$\sim$F$^2$(q$^2$)), the distribution at threshold is
unexpectedly isotropic. Here, F$^2$(q$^2$) is the $^3$He form factor
depending on the momentum transfer q. The observations are consistent with the 
assumption of an $\eta$-mesic nucleus which isotropically decays into the
coherent $\eta$ channel.\\
There are different theoretical efforts for describing the experimental
data. A plane 
wave impulse approximation (PWIA) calculation by Kamalov and Tiator
\cite{kamalov,tiator} does not deal with final state interaction (FSI) effects
(dotted line in fig. \ref{cs_coh}). FSI are
explicitly taken into account in the 
calculations of Shevchenko and collaborators \cite{shev1s,shev2s}
The model
employs the finite rank approximation (FRA) which excludes virtual
excitations of the nucleus during the interaction with the $\eta$ meson. 
The calculation is performed in such a way as to reproduce the real part of the
scattering length to be 0.75 fm (dashed and
dot-dashed lines in fig. \ref{cs_coh}). 
A third calculation has been performed by Fix and Arenh{\"o}vel 
\cite{fix_co3}. In contrast to the assumptions of Shevchenko which neglect
target excitations, a two-body quasi particle separation is applied to the
$\eta$-3N problem. The analysis obtains values for the
$\eta$-$^3$He 
scattering length that do not allow a bound state of $\eta$ and nucleus.
However, the existence of a virtual pole near threshold is claimed - implying
a quasi-resonance with positive binding energy that modifies the threshold
behaviour of the coherent cross section. The cross section shows a sharp
cusp effect at threshold but stays a factor of 2 below the measured data
(solid line in fig. \ref{cs_coh}).\\
Earlier measurements of the $\eta$ production off $^3$He \cite{mayer,bilger} in a different
initial state - the proton induced reaction on the deuteron - showed a
similar energy dependence when dividing the different phase space factors
out and thus comparing the average amplitude as a function of the $\eta$
momentum in the CM frame (fig. \ref{saclay}). Only the two data points
between 0.6 and 0.8 fm$^{-1}$ seem to be significantly lower in the
photoproduction data compared to the proton induced case. Also shown in
fig. \ref{saclay} is an optical model fit to the data below 0.5
fm$^{-1}$.\\ 
Additional information about a possible $\eta$-mesic state can be deduced by
studying another decay channel for an $\eta$-mesic nucleus. The quasi-bound
system will decay into $\eta$$^3$He for 
energies above the $\eta$ production threshold. Since
the phase space is limited, the momentum of 
the decaying $\eta$ meson will preferentially be low and hence there is a high
probability for the $\eta$ to be captured by a nucleon into an
S$_{11}$(1535) resonance. This resonance has some 50\% branching ratio into
$\eta$N but also roughly the same probability to decay into $\pi$N. The
advantage of the latter decay mode is that the $\eta$ meson is acting as
an intermediate particle only. Only the lighter $\pi$ meson
is emitted in the final state. Hence the reaction can also take place below the $\eta$
production threshold. Below threshold, the decay into $\eta$$^3$He is not
possible and the decay proceeds exclusively into the $\pi$N channel.\\
\begin{figure}
   \includegraphics[width=\columnwidth]{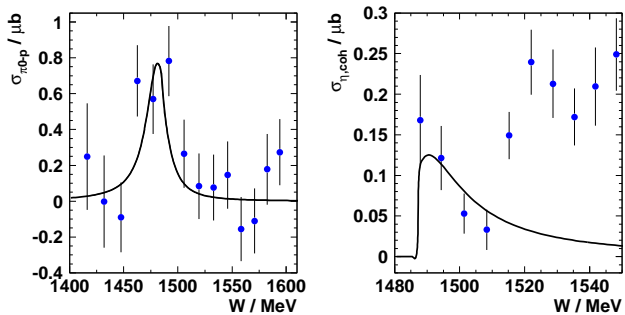}
  \caption[Comparison]{Comparison of the decay channels of an
$\eta$-mesic nucleus together with a simultaneous fit. The resonance
position is at 1481.2 MeV $\pm$ 4.2 MeV with a full width of (25.3 $\pm$ 6.1) MeV.
\label{comparison}}
\end{figure}
The signature for the decay of an $\eta$-mesic nucleus into $\pi$N is given
by correlated pairs of $\pi^0$ and p. With a low momentum $\eta$
in the intermediate state, the decaying S$_{11}$ will also be at low
momentum. Thus, the relative angle of $\pi^0$ and proton has to be
near 180$^\circ$ in the $\gamma$-$^3$He CM system. The triangles in the
left panel of 
fig. \ref{mesic} show the excitation function for relative angles between
170$^\circ$ and 180$^\circ$ as a function of
$W$=$\sqrt{2\cdot{}E_\gamma\cdot{}m_{^3He}+m^2_{^3He}}-m_d-B_{^3He}$. Here, the total
CM energy is reduced by the mass $m_d$ of the residual nucleus and the $^3$He
binding energy of $B_{^3He}$= -5.5 MeV (the $\eta$-$^3He$ binding energy for a
structure at resonance mass $W_R$ would then be given as
$B_\eta=W_R-(m_\eta+m_p)$).
The reaction channel is
dominated by the quasifree $\pi^0$ production. This background
channel is determined by
subtracting the scaled $\pi^0$p excitation function for relative opening
angles of 150$^\circ$ to 170$^\circ$.
A remaining structure just below the production threshold for 
$\eta$ mesons may indicate the formation for an $\eta$-mesic
$^3$He nucleus. Its peak position is $W_R$=1483 MeV ($\pm$ 5 MeV) with a
width $\Gamma$ = 39 MeV ($\pm$ 21 MeV). The
experimental resolution is given by the tagger spectrometer resolution of 
about 2 MeV. The statistical significance of the structure
$\sigma=\frac{S}{\sqrt{(\Delta S)^2+(\Delta BG)^2}}$ (S is the number of
signal counts and BG the corresponding background under the peak) is
$\sim$3.5$\sigma$. 
fig. \ref{comparison} shows a comparison of both
decay channels with a simultaneous fit of both channels. The result,
$W_R$=1481.2
MeV ($\pm$ 4.2 MeV) and $\Gamma$ = 25.3
MeV ($\pm$ 6.1 MeV), is consistent with
the results given in fig. \ref{mesic} but reduces the error bars on the
extracted parameters. The different phase space 
factors for the two decay channels
were taken into account by assuming a
single intermediate S$_{11}$(1535) resonance neglecting multiscattering
processes. 
The fit function is a Breit-Wigner distribution
$f_{BW}(W)=\frac{\frac{1}{4}\Gamma^2}{(W-W_R)^2+\frac{1}{4}\Gamma^2}$ folded
with the corresponding phase space factors
\vspace*{-0.25cm}
\begin{center}
\begin{math}
\nonumber
B_{\pi,\eta}(W)=b_{\pi,\eta} \cdot
\sqrt{\frac{(W^2-(m_p-m_{\pi,\eta})^2) \cdot 
(W^2-(m_p+m_{\pi,\eta})^2)}{4{}W^2}}.
\end{math}
\end{center}
\vspace*{-0.25cm}
Here, $b_{\pi,\eta}$ are
constants that fix the partial widths for the decay of the 
$S_{11}$ into $\pi$ and $\eta$ to 75 MeV at the pole mass of 1535 MeV,
respectively. Further, $m_p$ is the proton mass and $m_{\pi,\eta}$ are the
meson masses. \vspace*{0.2cm}\\
In summary, we have investigated the photoproduction of $\eta$-mesic $^3$He
by studying two possible decay channels - the coherent $\eta$ production and
the decay into $\pi^0$p. Both decay channels show indications for
photoproduction of $\eta$-mesic $^3$He. In the $\eta$ channel, a
resonance-like behaviour of the coherent cross section associated with an
isotropic angular distribution in the threshold regime is observed. In
contrast, a strong forward peaking is expected for 
coherent $\eta$ production from form
factor considerations. In the $\pi^0$p 
channel, correlated $\pi^0$-proton pairs with relative angles near 180$^\circ$
in the CM system have been observed that give rise to a peak-like structure at
energies slightly below the $\eta$ production threshold. The extracted
resonance 
parameters (binding energy=(-4.4 $\pm$ 4.2) MeV, full width=(25.6 $\pm$ 6.1) MeV) are
consistent with expectations for $\eta$-mesic nuclei.\\
It is a pleasure to acknowledge inspiring discussions with W. Cassing
and C. Wilkin. We thank the accelerator group of the Mainz Microtron
MAMI, as well as the other technicians and scientists of the Institut
f{\"u}r Kernphysik at the Universit{\"a}t Mainz for the outstanding support.
This work was supported by DFG SPP 2034, SFB 221, SFB 443, the U.K.
Engineering and Physical Sciences Research Council, and Schweizerischer
Nationalfonds.\\
\bibliography{./mybibliography}
\end{document}